\documentstyle[aps,multicol,pre]{revtex}
\title{On a mechanism low-pressure insertion of chain molecules\\ 
into crystalline  matrices.}

\author{E.V. Vakarin, J.P. Badiali   
}

\address{
 LECA ENSCP-UPMC-CNRS, 11 rue P. et M. Curie,
75231 Cedex 05, Paris, France
}

\begin{document}

\maketitle

\begin{abstract}
A microscopic mechanism of  low-pressure insertion and separation
of chain-like molecules in host matrices is proposed. It is shown that the 
intramolecular correlations combined to appropriate host activities are
responsible for a low-pressure condensation of chain molecules. This allows
recover a fine structure of the isotherms and
to explain recent experiments on the insertion of $C_2H_2$ and $CO_2$
guest species. We argue that the mechanism should be dominant in 
low-dimensional host geometries, where the entropic effects are strongly 
suppressed and the major factors are the chain connectivity and packing.

 \end{abstract}

\begin{multicols}{2}
\section{Introduction}
Insertion of guest species into host matrices has recently attracted a
considerable interest as a designing tool for hydrogen storage 
systems\cite{Zuttel}, rechargeable batteries and superconducting
devices\cite{Slusky}. Recent progress in synthesis of new composite
materials\cite{Yaghi,Ferey} opens novel perspectives in separation and 
storage of larger molecules (e.g., acetylene, methane or even longer 
hydrocarbons) which have important technological applications. In 
particular, a highly controlled acetylene accommodation in a metal-organic 
microporous material has recently been reported\cite{Matsuda}. It has been 
found that due to formation of regular hydrogen bonds the guest molecules 
are arranged into dense one-dimensional strips at very 
low relative pressures. This allows for a safe storage at a density much 
higher than the compression limit of free acetylene. In contrast, $CO_2$ 
isotherms exhibit a much weaker adsorption in this pressure domain. This 
offers an efficient way of separation of these two compounds. 
In this context it has been 
emphasized\cite{FereyN} that the fabrication of high-capacity adsorbents, 
sensitive to particular  guest species, requires a better knowledge of the 
host-guest and guest-guest interactions. In particular, our understanding of 
a microscopic mechanism, that allows to reach high densities inside a host 
marix at very low bulk pressures,  would be quite helpful for a designing of 
novel storage and separation systems.

In this Letter we report on a possible microscopic mechanism, responsible
for a remarkable enhancement in the insertion of chain-like molecules
into crystalline host matrices at low bulk pressures. 
The main idea is to note that in the host-guest contact
the guest acts as a thermodynamic subsystem that  
has an impact (energetic and entropic) from
its bulk. This induces a correlation for the guest species, such
that their accommodation is not driven only by the accessibility of the 
matrix space but also by effective (density-dependent) interactions, arising
from the packing and the chain connectivity effects. For planar geometries
this mechanism has already found experimental 
evidences\cite{voronov,voronov1} and interesting kinetic 
interpretations\cite{minko}.  
%%%%%%%%%%%%%%%%%%%%%%%%%%%%
\section{Insertion model}
%%%%%%%%%%%%%%%%%%%%%%%%%%%
The host is modeled
as a lattice of adsorbing sites with the lattice spacing $d$ and 
coordination number $q$. The lattice is assumed to be absolutely rigid, such 
that possible guest-induced distortions\cite{PRBint} or 
restructurings\cite{JPCB} are excluded. The guest molecule is modeled as a 
chain (length $m$) of hard-sphere segments\cite{TPT} (diameter $\sigma$). In 
this way we take into account the chain connectivity and the excluded volume 
effects.      Therefore, the guest fluid is characterized by its bulk 
density $\rho$ (or packing $\eta=\pi\rho\sigma^3/6$) and  the chain length 
$m$. The latter is the number of hard sphere segments which mimic the 
functional groups (e.g. $m=2$ for $C_2H_2$). The bulk equation of state can 
be approximated as\cite{TPT} \begin{equation}
\label{P}
\frac{\beta P}{\rho}=\frac{\beta P_{hs}}{\rho}-
\frac{m-1}{m}
\left(
1-\frac{\eta}{2-\eta}+\frac{3\eta}{1-\eta}
\right)
\end{equation} 
where $P_{hs}$ is the hard sphere contribution and $\beta=1/(kT)$ is the
inverse temperature. In what follows this equation of state will be used to 
make a link between the bulk density $\rho$ and the pressure $P$. 
     
The adsorption potential 
$U_s({\bf r}_i)$ is given by the sticky site model\cite{bad,sperc}
\begin{equation}
\exp(-\beta U_s({\bf r}_i))=
1+\lambda \sum_{{\bf R}_m} \delta({\bf r}_i-{\bf R}_m)
\label{sticky}
\end{equation}
where ${\bf r}_i$ is a chain segment position,
${\bf R}_m$ is a position of a lattice site and $\lambda$ is
the stickiness parameter. The latter can be related to the segment
adsorption energy $\epsilon$ through $\lambda=\exp(\beta\epsilon)$. 
This singular one-body potential allows one
to perform an exact integration in the partition function  
\begin{equation}
Z=Z_{ref}\sum_{n=0}\frac{\lambda^n}{n!} \sum_{\{{\bf R}_k\}} 
\rho_n^{ref}({\bf R}_1,...,{\bf R}_n)
\label{Z}
\end{equation}
where $Z_{ref}$ is the partition function for the same system but
without the adsorbing potential, and 
$\rho_n^{ref}({\bf R}_1,...,{\bf R}_n)$ is the $n$-body distribution 
function taken at positions of the lattice sites. In general, even in the 
absence of a specific adsorption, a periodic host structure induces a 
density redistribution and particle correlations around the lattice strands.
Nevertheless, in a low-pressure regime (on which focus), these effects can
be neglected and the reference state can be approximated by the bulk state. 
Therefore,  
we have in eq.~(\ref{Z}) an infinite series on $\lambda$ including the 
correlations of all orders for the reference state. If only pair correlations
are important then the problem can be mapped onto the lattice
gas model \cite{bad,sperc}
\begin{equation}
\Xi=Z/Z_{ref}=\sum_{\{t_i\}}\exp(-\beta H_{LG})
\end{equation}
with the Hamiltonian
\begin{equation}
H_{LG}=\sum_{ij}W({\bf R}_i,{\bf R}_j)t_it_j -
\sum_i\mu({\bf R}_i)t_i
\label{LG1}
\end{equation} 
in which  $\{t_i\}$ is a set of occupation numbers. The effective chemical 
potential $\mu({\bf R}_i)$ and pair interaction $W({\bf R}_i,{\bf R}_j)$
are closely connected with the properties of the fluid in the bulk phase.
Namely\cite{bad,sperc},
\begin{equation}
\beta \mu({\bf R}_i)=\ln(\lambda\rho_1^{ref}({\bf R}_i,\sigma/2)) 
\label{bmu}
\end{equation}
\begin{equation}
\beta W({\bf R}_i,{\bf R}_j)=-\ln(g_2^{ref}({\bf R}_i,{\bf R}_j))
\label{bW}
\end{equation}
where $\rho_1^{ref}({\bf R}_i)$ and $g_2^{ref}({\bf R}_i,{\bf R}_j)$ 
are respectively the one-body and 
pair correlation functions for the reference state (i.e., a state without 
specific adsorption). The latter is independent
of the adsorption site position because the lattice is translationally 
invariant. If $\sigma \approx d$ then we deal with the contact values of
the strand-segment density profile $\rho_1(\sigma/2)$ and the 
segment-segment correlation function $g_2(\sigma)$.  
The contact density is approximated by its hard wall-chain value\cite{sperc}
\begin{equation}
\rho_1(\sigma/2)=\rho \sigma^3
\left[
\frac{1+2\eta}{(1-\eta)^2}-\frac{m-1}{m}\frac{1}{1-\eta}
\right]
\label{rho}
\end{equation}
Here the second term describes a chain depletion compared to a purely
monomeric case ($m=1$). At low pressures (bulk densities) this effect 
is marginal and $\rho_1(\sigma/2)$ can be replaced by $\rho\sigma^3/m$ in 
this domain. In the Percus-Yevick approximation the pair correlation is 
given by\cite{Sandler} 
\begin{equation}
g_2(\sigma)=
\frac{1+\eta/2}{(1-\eta)^2}-\frac{m-1}{m}\frac{Q}{1-\eta}+
\left[\frac{m-1}{m} \right]^2 \frac{K}{\eta}
\label{g2}
\end{equation}
Here the first term corresponds to the hard sphere
packing effects which are dominating with increasing $\eta$. The second 
term describes a depletion with increasing $m$ due to the chain-chain 
repulsion - {\it the intermolecular correlation}. 
 The last term in $g_2(\sigma)$ is responsible for 
{\it the intramolecular correlation inside a chain}. The latter is 
dominating at low densities when the chain-chain contacts are 
rather rare and the segments correlate mainly through the chain 
connectivity. That is, a probability to find two randomly chosen segments
in the same chain is rather high. 
The singularity at $\eta \to 0$ is a flaw of the Percus-Yevick
approximation, which, however,  does not affect our conclusions.  
The coefficients $K$ and $Q$ control the 
corresponding magnitudes, which depend on the chemical nature of
molecules (e.g. chain flexibility) and on the dimensionality of
space (in 1D, for instance, the intermolecular correlations are realized
only through the chain ends). 
Note that the chain length $m$ is chosen as an 
external variable. This means that we do not consider the clustering process 
itself, assuming that the guest cluster composition remains fixed in the 
course of insertion. This assumption is valid if the adsorption is not 
dissociative or if the adsorption rate dominates the rate of dissociation in 
the fluid bulk.
%%%%%%%%%%%%%%%%%%%%%%%%%%%%%%%%%%%
\subsection{Effective interaction}
%%%%%%%%%%%%%%%%%%%%%%%%%%%%%%%%%
As we have already noticed, due to correlations in the reference state
(bulk), the adsorbed guest species experience an effective pairwise 
lateral interaction $\beta W=-\ln g_2(\sigma)$, where 
$g_2(\sigma)\approx g_2(d)$ is given by eq.~(\ref{g2}). 
In Figure~1 this quantity is analyzed as function of the packing $\eta$
at different chain lengths $m$. It is seen that $\beta W$ is a non-monotonic 
function. In a low-density (pressure) domain, dominated by the 
intramolecular correlations, the interaction is strongly attractive. This 
effect increases with the chain length. At intermediate densities the 
chain-chain competition for the available space (the intermolecular 
correlations) leads to an effective repulsion at neighboring lattice sites.
Finally, at high densities the chain connectivity plays only a marginal
role since a closely packed chain configuration becomes identical to
that of disconnected hard-sphere segments. This leads to an effective 
attraction\cite{bad,sperc}, similar to the depletion forces in colloidal
systems. Although our analysis is based on the approximate result
eq.~(\ref{g2}), our conclusions should hold in general because, 
from the experimental point of view\cite{voronov,voronov1,minko},
the density/concentration regimes (e.g. diluted, semidiluted and plateau 
regimes) are clearly distinguished. 
%%%%%%%%%%%%%%%%%%%%%%%%%%%%%%%%%
\subsection{Insertion isotherms}
%%%%%%%%%%%%%%%%%%%%%%%%%%%%%%%%%
Having determined the inputs to the lattice gas Hamiltonian
we can find the mean-field approximation for the free energy, which gives 
the adsorption isotherm\cite{bad,sperc}
\begin{equation}
\label{MF}
\theta=\frac{\lambda \rho_1(\sigma/2)[g_2(\sigma)]^{q\theta}}
{1+\lambda \rho_1(\sigma/2)[g_2(\sigma)]^{q\theta}}
\end{equation}
where $\theta=\langle \sum_i t_i \rangle/N$ is the coverage -
the guest density inside the matrix. Equation 
(\ref{MF}) has been solved numerically in order to determine the coverage
$\theta$ as a function of density $\eta$ and chain length size $m$.
Then a pressure dependence can be obtained through the equation of state
(\ref{P}). The result for a dimeric fluid ($m=2$) on a four-fold 
coordinated  lattice ($q=4$) is plotted in Figure~2. It is seen that 
the isotherm shape is coherent with the density (pressure) modulation of the 
effective interaction $\beta W$ (see Figure~1). Namely, at low pressures 
when the chain-chain contacts are negligible, the adsorption of one bead 
promotes the adsorption of the entire chain through the intramolecular 
correlations. As a result we have a remarkable increase of the guest density 
$\theta$ in a very narrow pressure domain.   This cooperative effect has 
been confirmed experimentally\cite{voronov}. At intermediate pressures the 
chains compete for the adsorption sites (the intermolecular correlations) 
and the isotherm exhibits a "slower" increase with a characteristic 
inflection point. The latter signals a cross-over to the packing regime, 
where the isotherm reaches a saturation. It is remarkable that the overall 
shape of the isotherm very closely resembles the one reported for 
acetylene\cite{Matsuda}, including the low-pressure feature (see the inset 
to Figure~1(a) in that paper) signalling a very efficient accommodation.
On the other hand, the $CO_2$ molecules (which are of
similar size as $C_2H_2$ and can also be viwed as diatomics) do not 
exhibit this effect. The chain length difference could explain the different 
adsorption behavior as a segregation phenomenon\cite{JCPseg}, even without 
considering a specific adsorption.

Therefore, we have to find out why the chains of the same length
could behave differently. With this purpose we have analyzed the parameters,
at which the low-pressure condensation occurs. We have studied the solution 
of eq.~(\ref{MF}) at $m=2$ and one-dimensional (1D) (q=2) geometry that 
mimics a channeled host structure, reported in \cite{Matsuda}. Interactions 
in the neighboring channels were neglected and the in-channel interactions 
were assumed to be governed by  $g_2(\sigma)$.   
 A fit to the experimental 
data\cite{Matsuda} is presented in Figure~3. It seen that the theory is
qualitatively consistent with the data. It is found that the low-pressure
feature sharpens with increasing stickiness $\lambda$ (segment binding 
energy). In order to be free from possible artifacts of the mean-field
approximation we have studied an exact solution for 
1D geometry\cite{Baxter}   
\begin{equation}
M=\frac{\sinh{(H)}}{\sqrt{\sinh^2{(H)}+e^{4J}}},
\end{equation} 
where $J=\beta W$, $H=\beta(\mu-W)/2$, $M=2\theta-1$.  
It is found that there are two regimes (see the inset in Figure~3). 
In a low-activity regime (small $\lambda$) the isotherm exhibits a gradual
adsorption without the low-density condensation, while in the high-activity
regime (large $\lambda$) we observe this feature. This can explain a 
striking difference between $C_2H_2$ and $CO_2$ \cite{Matsuda}. Although
the latter can also be viewed as diatomic (possibly with different inter- 
and intra-molecular constants $Q$ and $K$), its interaction with the host
is much weaker that that of $C_2H_2$, which is found\cite{Matsuda} to form 
two hydrogen bonds with non-coordinated oxygens in the host matrix.  

In conclusion, we have found that an interplay of the inter- and 
intramolecular correlations combined to a variation in the host activity
is an efficient tool for low-pressure insertion and separation of chain
molecules. The proposed mechanism should be dominant in low-dimensional
geometries, where the entropic effects (due to the chain conformation and 
reorientation) are strongly suppressed. We argue that the mechanism is not
exclusive to particular guest molecules, such that the $C_2H_2$ insertion
should be considered as a prototypic example. Therefore, a detailed study 
of a correlation between the host geometry, activity and the chain length 
and architecture could open novel perspectives in the storage and separation 
of other technologically important species. In particular, a proper control 
of the guest-induced matrix distortions\cite{PRBint} or 
restructuring\cite{JPCB} could lead to a suitable matching or 
mismatching between the host geometry and the guest architecture. This
offers a promising tool for swithching between the high- and low-activity
regimes, increasing the insertion selectivity in a given domain of
controlling thermodynamic parameters (e. g. pressure, density, 
 concentration).  

%%%%%%%%%%%%%%%%%%%%%%%%%%%%%%%%

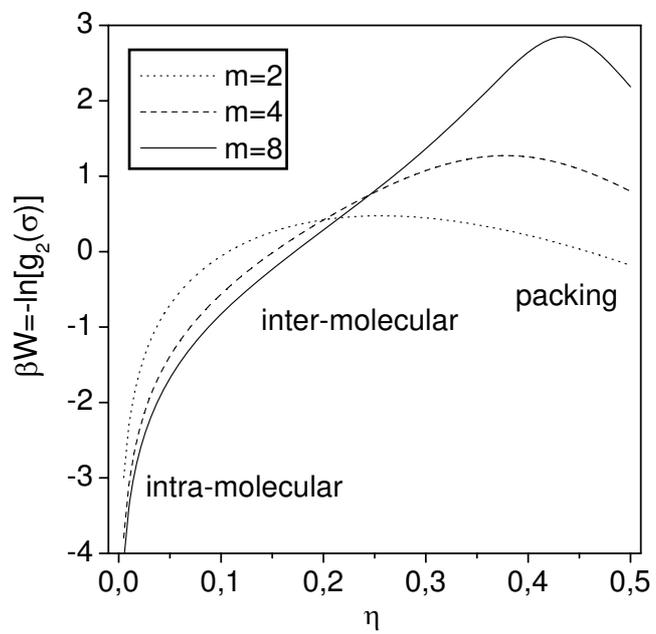
\begin{figure}
\caption{ The effective lateral interaction $\beta W$ as a function
of the bulk density $\eta$ at different chain lengths $m$.  
} 
\end{figure}

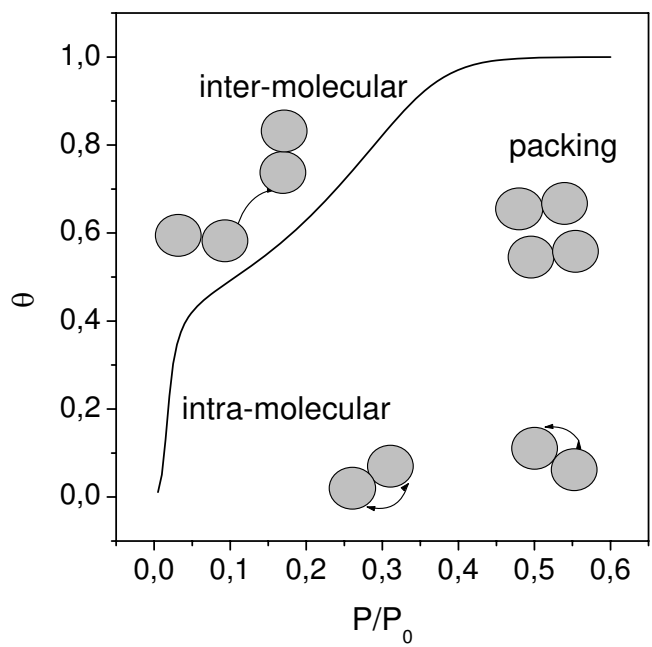
\begin{figure}
\caption{Mean-field adsorption isotherm for dimers $m=2$ on a lattice with
coordination $q=4$. Other parameters are $\lambda=800$, $K=0.15$, $Q=2$.
The saturation pressure $P_0$ corresponds to $\theta \to 1$. The inset
shows typical chain arrangements and correlations in different pressure 
domains.} 
\end{figure}

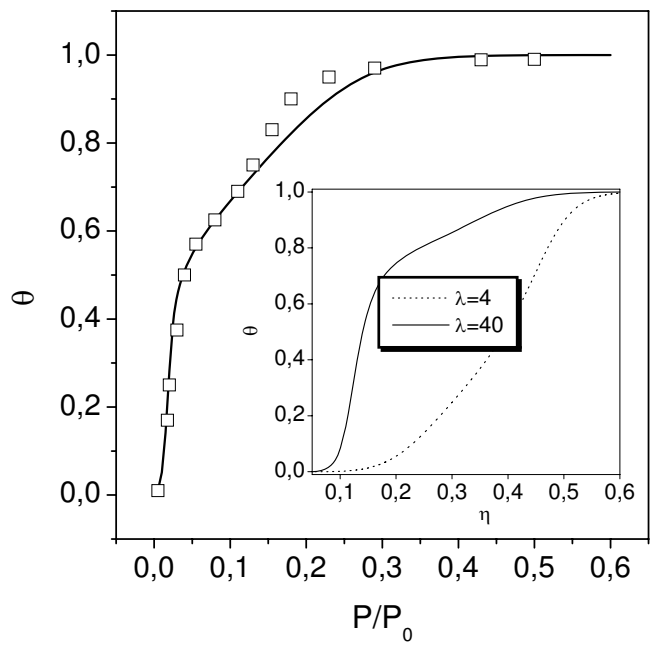
\begin{figure}
\caption{ Fitting of the mean-field isotherm (line) to the experimental
data of Ref.[5] (symbols) for $C_2H_2$ insertion. The saturation pressure
$P_0$ was taken to be $5 kPa$. 
The inset shows the
exact 1D isotherms at different activity regimes (see text).    
} 
\end{figure}

\end{multicols}
\end{document}